# RPCA-Based High Resolution Through-the-Wall Human Motion Detection and Classification

Qiang An*, Shuoguang Wang*, Wenji Zhang, Hao Lv, Jianqi Wang, Shiyong Li†, and Ahmad Hoorfar†, *IEEE Senior Member*

*Abstract*—Radar based assisted living has received great amount of research interest in recent years. By employing the micro-Doppler features of indoor human motions, accurate recognition and classification of different types of movements become possible. Whereas, most of the existing works are focused only on free space detection, the literature on detection and recognition of human motions in through-the-wall scenarios is still in its infancy. As can be anticipated, the wall media and indoor static non-human targets would cause clutters and significantly corrupt the motion information of human subjects behind wall. However, no relevant work is reported to effectively handle this problem. In the present work, we aim to fill the gap and propose to use a low center-frequency ultra-wideband (UWB) radar system to probe the behind wall scene. Then, a Robust Principal Component Analysis (RPCA) based subspace decomposition technique, as its first reported implementation, is employed not only to remove the stationary clutters in raw range slow-time map but also to mitigate the multipath effects in the time-frequency map. Onsite experiments of detecting human motions behind a single layer of concrete wall is carried out to investigate the performance of the technique. Lastly, a two dimensional (2D)-PCA algorithm-based motion classification is provided to further verify the effectiveness of the proposed technique. Classification result shows that an enhanced recognition capability can be achieved using the proposed technique in detection and classification of indoor human motions.

*Key words*---Assisted living, micro-Doppler signatures, robust principal component analysis (RPCA), 2D-PCA, high resolution range profile (HRRP), range slow-time map, multipath effects, range-max time-frequency representation (R-max TFR)

## I. Introduction

The emerging concept of monitoring indoor human activities has lately resulted in intensive studies to meet the increasing demands for improving the living quality and intelligent caring of the elderly as various societies around the wrold are rapidly facing an aging population [1]. As compared to other sensing approaches such as sound, thermal and video, radar has found its prominent application advantages in this area. Not only can the electromagnetic (EM) wave radiated by radar antenna penetrate through a certain thickness of lossy dielectric media to remotely retrieve the target information of interest, but it can also protect the privacy of the subject under monitoring [1]. Inspired by the pioneering work of V.C. Chen, [3] who introduced time-frequency analysis to describe different human body parts micro-motions in radar observation, numerous research works for radar-based indoor monitoring have been reported [1, 4, 5].

Whereas, most of these work considered only the free space scenario, which is mainly motivated by the fact that the construction material of partition walls in a room is mostly drywall in the US. Thus, to reach the highest possible Doppler resolution, while maintaining an acceptable hardware cost, the radar systems adopted in these works often operate around millimeter wave band. However, the situation is not ubiquitous in other regions of the world. Types of walls like concrete walls and cinder block walls are typically present within tall residential buildings in densely populated urban areas. In order to penetrate through these types of wall media, a low operating center-frequency around 1-3 GHz is often suggested [2]. Nevertheless, the EM wave would be strongly attenuated and distorted because of the impact of reflection, refraction and dispersion when it penetrates through lossy walls [30, 31, 32]. As a result, the captured micro-Doppler signatures of behind wall human subjects will become very weak in intensity. Meanwhile, strong DC components will be introduced due to the constant phase offset generated in the wall media. In fact, the antenna coupling and the multipath effects caused by stationary objects in the investigation region both contribute to the constant phase term. The authors in [5] and [6] utilized a narrow band radar system with the center-frequency fixed at 5.8 GHz and 6.5 GHz, respectively, to study the micro-Doppler signatures of behind-the-wall human motions, in which only a small set of motions, namely, stationary non-periodic movements, can be recognized. As micro-Doppler features of other motions such as stationary periodic movements and forward-moving movements are very similar in their respective subsets, one is practically unable to be discriminate, let alone classify and identify such movements.

The very same problem exists in millimeter wave radar based indoor human motions detection application. B. Erol in [7]

Qiang An* and Shuoguang Wang* contribute equally to this paper. Shiyong Li† and Ahmad Hoorfar† are the Co-corresponding authors for this paper.
Qiang An, Hao Lv, and Jiangqi Wang are with the department of Biomedical Engineering, Fourth Military Medical University, Xi'an 710032, China (qan01@villanova.edu, fmmulvhao@fmmu.edu.cn, wangjq@fmmu.edu.cn).
Ahmad Hoorfar and Wenji Zhang are with the Antenna Research Laboratory, Center for Advanced Communications, Villanova University, Villanova, PA 19085 USA (ahoorfar@villanova.edu, wenjizhang@gmail.com).
Shuoguang Wang and Shiyong Li are with the Beijing Key Laboratory of Millimeter Wave and Terahertz Technology, Beijing Institute of Technology, Beijing 100081, China (shuoguangwang@outlook.com, lisy_98@bit.edu.cn).

proposed to use an ultra-wideband (UWB) radar system to address the problem, in which an optional feature representation domain, range over slow-time, is exploited to observe the range variation information over time for different types of motions with high resolution. The motions which might smear in the time-frequency feature representation domain collected using a narrow band radar system, can then be easily discriminated in the range slow-time domain, or the so called the range map [8].

Based on the aforementioned works, a low center-frequency UWB radar system turns out to be a good choose for studying the characteristics of behind-the-wall indoor human motions. However, due to restriction of Federal Communication Commission (FCC) regulations in the US, of which unlicensed usage of UWB systems is restricted to frequency range from 3.1 GHz to 10.6 GHz [1], no results concerning to through wall indoor human motions detection are reported for L (1-2 GHz) and S (3-4 GHz) bands, i.e., the bands the EM wave presents a better penetration ability than the frequency bands used in [5] and [6].

In this paper, a stepped frequency continuous wave (SFCW) UWB multiple-input multiple-output (MIMO) radar system with an operating frequency ranging from 400 MHz to 4.4 GHz is adopted to realize through wall indoor human motions detection. The corresponding wideband micro-Doppler signatures of behind wall human motions is calculated using the method proposed in [9], in which the target range bins indicating motions of interest are first coherently added together, then the traditional time-frequency analysis approach, short-time Fourier transform (STFT), is applied to process the resulting one dimensional (1-D) slow time data.

However, as can be expected, the obtained raw range map would be characterized by strong DC clutter components, multipath components and noise components, which are introduced by wall media and static non-human objects in a rich indoor environment. Thus, before micro-Doppler features is extracted, these effects should be first removed. Traditionally, mean value subtraction approach is employed to suppress the stationary DC clutters in the range map [10], [11]. Because the positions of wall and static indoor objects don't change along with the extension of sampling process, they can be removed effectively using the above approach. The approach essentially performs similar to the wall clutter removal scheme proposed in [12], in which only zero-frequency component is suppressed by spatial filtering. However, the latter effort is unable to remove the multipath and noise components, and at the same time, it distorts the phase information of human subject movement information since it is essentially a nonlinear filter.

In addition to the above approach, recently a state-of-art Robust Principle Component Analysis (RPCA) based subspace decomposition approach has proven its effectiveness in mitigating stationary clutters [13-17]. It aims to recover a low-rank matrix (stationary background with only a limited number of independent rows) and a sparse matrix from highly corrupted measurement concurrently. The approach was first reported to remove the outliers in computer vision [13]. Later, it was applied in through wall radar imaging (TWRI) to mitigate wall reflections [14], suppress in-wall clutters [15] and detect antipersonnel mines in ground penetrating radar (GPR) imaging [16]. When it comes to through-the-wall indoor human motion detection, from a RPCA perspective, the stationary DC clutters and noise components in the raw range map can be modeled as low-rank component, while human moving information can be modeled as sparse component [17]. Then, it can be inferred that the information of behind-wall human motions can be accurately extracted from a highly corrupted range map with high confidence using the RPCA method, which would definitely increase the identifying characteristics of behind wall indoor human motions in the time-frequency map. We note that an implementation of standard PCA for TWRI was first reported in [33] and was shown effective in removal of wall and other ambient clutters using polarimetric radar.

This paper marks the first attempt to use RPCA to mitigate the stationary DC clutters and noises in the raw range map for through-the-wall human motion problem under study. The use of RPCA for the present problem, has the following key advantages. First, by using the Augmented Lagrange Multiplier (ALM) algorithm in [17], RPCA requires no tuning of parameters. Second, the required workload of RPCA method is comparable to that of mean-value subtraction method, which means real-time processing for the through-the-wall human objects is achievable using the proposed method. Third, allowing for the phase information preserving capability of the RPCA method, accurate micro-Doppler features of behind-wall motions can be retrieved. In this work, we collected the real data in laboratory experiments to validate the proposed method. It should be noted that the spectrograms in this paper are all calculated using the novel range-max time-frequency representation (R-max TFR) method that we proposed in [34], in which we proved that R-max TFR approach outperforms all other reported time-frequency analysis methods and feature enhanced spectrograms can be obtained using the method.

The paper is organized as follows. In Section 2, the signal model for the SFCW ultra-wideband MIMO radar system is detailed and the time-frequency analysis method is presented. In Section 3, the RPCA method as applied to the present through-the-wall problem is discussed. In Section 4, experimental demonstrations and classification results for different behind-wall human motions are presented. In Section 5, the concluding remarks are provided.

## II. THOUGH WALL INDOOR HUMAN MOTIONS MICRO-DOPPLER ANALYSIS

### A. Signal model for SFCW MIMO radar system

A SFCW UWB MIMO radar system, operating in a frequency consisting of $N$ frequency bins with a frequency step of $\Delta f$, is adopted to detect the through wall micro-Doppler signatures of indoor human motions. The antenna array consists of $N_t$ transmitting and $N_r$ receiving UWB antennas. A schematic diagram of measurement configuration is depicted as below in Fig. 1.

While the analyses in this and the following sections are general, we have implemented the proposed technique, using both numerical simulations and laboratory experiments, to a MIMO radar with $N_t = 2$, $N_r = 4$, operating frequency band of 40 MHz to 4.4 GHz with a frequency step of 5 MHz, and a pulse repetition frequency (PRF) of 113 Hz. The resulting system has a range resolution about 3.4 cm and is thus capable of providing both high resolution range and Doppler information.

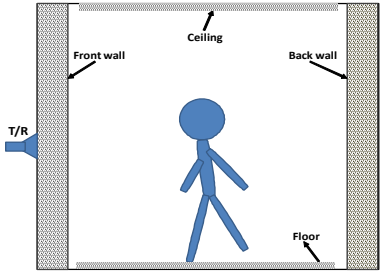

Figure 1: Schematic of through wall indoor human motion measurement

Assuming the emitted SFCW UWB signal contains a total of $N$ frequency bins, the transmitted signal for the $n-th$ sweeping frequency can be written as below,

$$s_t(n,t) = rect\left(\frac{t}{T}\right) exp\left(j2\pi(f_0 + n\Delta f)t\right) \quad (1)$$

where $rect$ is a rectangular window function, $T$ is the modulation period, $f_0$ is the starting frequency, $\Delta f$ is the frequency increment.

The received signal, which contains valuable information about modulated Doppler features of behind-the-wall and indoor human subject presenting different types of movements at distance $d$, can then be expressed as,

$$s_r(n,t) = a_d \cdot rect\left(\frac{t-t_d}{T}\right) exp\left[j2\pi(f_0 + n\Delta f)(t-t_d)\right] \quad (2)$$

Where $t_d$ denotes various time-delays backscattered from front and back walls, static indoor objects and various body parts of the human target in motion; $a_d$ is the respective reflectivity intensity corresponding to a given time-delay $t_d$.

After frequency mixing, low-pass filtering and digital sampling, the received signal can be organized into a two-dimensional (2-D) matrix. Each column of the matrix, termed high resolution range profile (HRRP), is obtained by applying the inverse Discrete Fourier transform (IDFT) to the received signal collected at all frequency bins, while the samples in each row correspond to different probing signals in slow-time. The above 2-D matrix is known as the range map, in which the range over slow-time variation of behind-the-wall human motions is incorporated. It can be mathematically written as,

$$s(m,t) = IDFT_M(s_r(n,t)), \quad m = 1, 2, \cdots M \quad (3)$$

where $M$ denotes the next power of 2 higher than $N$.

The range maps for two types of behind-the-wall human motions, namely sitting and crawling toward the wall, which are obtained using the above SFCW UWB MIMO radar system, are shown in Fig. 2-(a1) and Fig. 2-(b1) respectively. As shown in these figures, strong stationary DC clutters, introduced by the wall and static indoor objects, have completely obscured the target movement information in both cases.

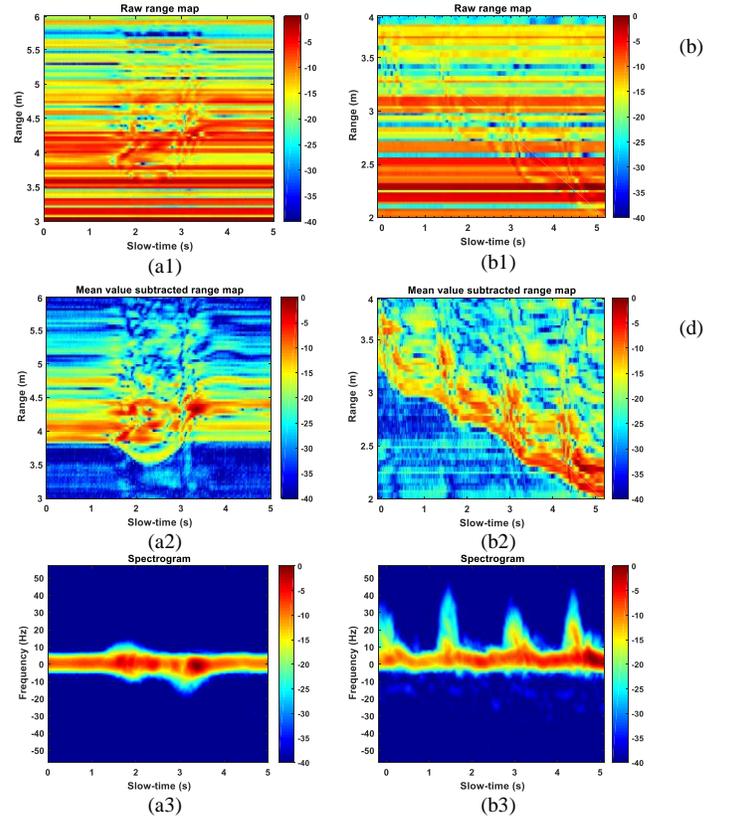

Figure 2: Raw range map, mean value subtraction results, and corresponding spectrogram for two types of behind-the-wall motions. (a1) Range map for sitting. (b1) Range map for forward crawling. (a2) Mean value filtered result for (a). (b2) Mean value filtered result for (b). (a3) Spectrogram of (a2). (b3) Spectrogram of (b2).

In order to remove such clutters, a mean value subtraction-based approach along the slow-time dimension is applied [10], [11]. Fig. 2-(a2) and Fig. 2-(b2) show the clutter suppression results using this approach. The stationary DC clutters are suppressed to a certain extent and the trajectories of target movement become apparent. However, it is observed that significant noise components fill the filtered range map. At the same time, additional DC clutters are introduced at locations where abrupt variation of motions occurred in slow-time dimension. This is another disadvantage of the approach. Thus, a new approach should be devised to cope with this problem.

Before we go much further, let us first describe how one can evaluate the time-frequency representation for the above motions.

B. *Micro-Doppler analysis for behind wall human motions*

Micro-Doppler analysis is proposed to depict the features of a target in motion in time-frequency domain [3]. Rotating, oscillating or vibrating parts of the target will introduce additional frequency modulations centered around main Doppler shift, caused by translational motion, in time-frequency map. The information then serves as characteristic features to realize the detection, recognition and classification of a specific target.

Different from the traditional time-frequency analysis for narrow-band signal using STFT, an extra procedure is needed to obtain the micro-Doppler signatures of wideband signal, in which desired range bins in mitigated stationary DC clutters' removed range map should be first coherently summed together.

The range-stacked slow-time data is calculated as below,

$$s(t) = \sum_{m=M_1}^{M_2} s(m,t) \quad (4)$$

where $M_1$ is the starting index of regions of desired range bins and $M_2$ is its ending index.

Then, the spectrogram of behind-the-wall indoor human motions can be readily obtained by applying STFT to the above range-stacked slow-time data, resulting in

$$SPEC(n,k) = \left| \sum_{t=0}^{K-1} s(t) h(t-n) \exp(-j2\pi kt/K) \right|^2 \quad (5)$$

where $h(t)$ is the window function, specifically a Hanning window in this paper. In order to strike a balance between frequency resolution and time resolution, the length of the Hanning window is selected to be 32 with an overlap of 31.

Fig. 2-(a3) and Fig. 2-(b3) show the spectrograms evaluated using the approach detailed in this part. As expected, a strong DC component appears in micro-Doppler signatures around zero-frequency, which corresponds to the constant phase offset. Also, additional DC clutters are introduced in the mean value subtraction filtered range map. These will attenuate the already very weak micro-Doppler features of behind-the-wall motions and thus make it difficult to realize motion recognition and classification.

### III. RPCA Based Through Wall Indoor human Motion Micro-Doppler analysis

Low-rank and sparse representation (LRSR) based subspace segmentation model declares that a given observation $D \in C^{m \times n}$ can be decomposed into a low rank matrix $L \in C^{m \times n}$ and a sparse matrix $S \in C^{m \times n}$.

In through-the-wall micro-Doppler analysis context, the wall is typically modeled as a laterally homogeneous layer of media, whose reflections constitute the stationary DC components in raw range map [18]. Other static objects, such as desks and chair in a densely populated indoor environment, produce similar DC components. At the same time, the reflected signals from body parts of behind-the-wall human in motion, which is changing constantly along the observation time, compose the non-stationary components of the range map.

Based on above description, by converting the through wall human motion radar observation into a LRSR decomposition problem, the following pairwise relationships can be established. The range map is equivalent to the observation matrix $D$. The stationary DC clutters is corresponding to the low-rank component $L$. The non-stationary component capturing the human motion information is equivalent to the sparse component $S$.

In order to recover the low-rank component $L$ and the sparse component $S$ in the LRSR model, the RPCA method is proposed here to overcome the above challenges for behind-the-wall human motion detection and classification. The LRSR decomposition problem can be formulated as below,

$$\arg\min_{L,S} \|L\|_* + \lambda \|S\|_1 \quad \text{s.t.} \quad D = L + S \quad (6)$$

Mathematical analysis has shown that when the value of the tuning parameter is set to $\lambda = 1/\sqrt{max(m,n)}$, perfect recovery of both components can be achieved [17]. Thus, although fine tuning of $\lambda$ can improve the recovery performance for different practical problems, we use the empirical value in this work for generalization purpose. Here, $m$, $n$ refers to the number of rows and columns of matrix $D$ respectively.

The ALM algorithm is utilized in this paper to solve the above LRSR problem [19], [20]. Not only does the algorithm work stably without parameter tuning, but it is also computationally efficient since the number of iterations is bounded by $rank(L)$ throughout the optimizing procedure.

By employing the RPCA method, the recovered sparse components of the results in Fig. 2-(a1) and Fig. 2-(b1) are depicted in Fig. 3-(a1) and Fig. 3-(b1), respectively. When compared to mean value filtered results in Fig. 2-(a2) and Fig. 2-(b2), a clean range map with detailed human subject body parts movement information is highlighted. Not only the stationary DC clutter components are removed thoroughly, the noise components are also significantly suppressed. Fig. 3-(a2) and Fig. 3-(b2) show the corresponding spectrograms of Fig. 3-(a1) and Fig. 3-(b1), in which zero-frequency DC components are mitigated completely as compared to the results in Fig. 2-(a3) and Fig. 2-(b3). These results indicate that the multipath components are also mitigated by using RPCA method. At the same time, much more evident micro-Doppler features are observed in Fig. 3-(a2) and Fig. 3-(b2), which will certainly improve the recognition and classification accuracy of behind-the-wall human motions.

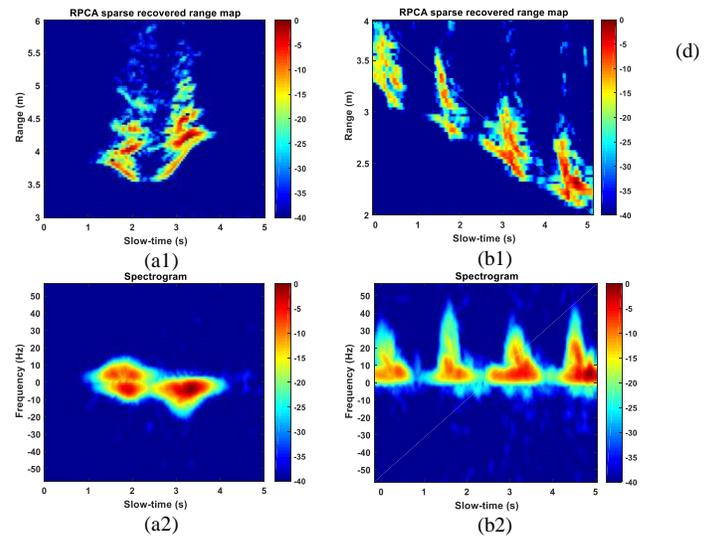

Figure 3: RPCA decomposition results, and corresponding spectrogram for two types of behind wall motions. (a1) Sparse motion component for sitting. (b1) Sparse motion component for forward crawling. (a2) Spectrogram of (a1). (b2) Spectrogram of (b1).

Since different motions are obviously characterized in the above results, both RPCA decomposed sparse range map and its corresponding feature enhanced spectrogram can be utilized to perform motion recognition and classification.

## IV. Experiment Results

In order to validate the effectiveness of the proposed RPCA-based technique in clutter mitigation, noise suppression and micro-Doppler feature enhancement, a series of through-the-wall human motion detection experiments were carried out with the results outlined in this section. The experimental scene is depicted as shown in Fig. 4. A human subject stands behind a fly ash concrete block wall performing

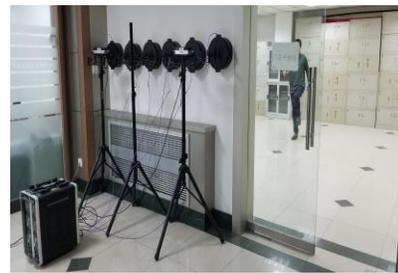

Figure 4: Through wall measurement scenario for motion detection.

different types of motions. The wall is of a thickness of 30 cm. The MIMO antenna array consists of 2 transmitting and 4 receiving planar logarithmic spiral antennas. As shown in Fig. 4, the array is placed against the exterior surface of the wall

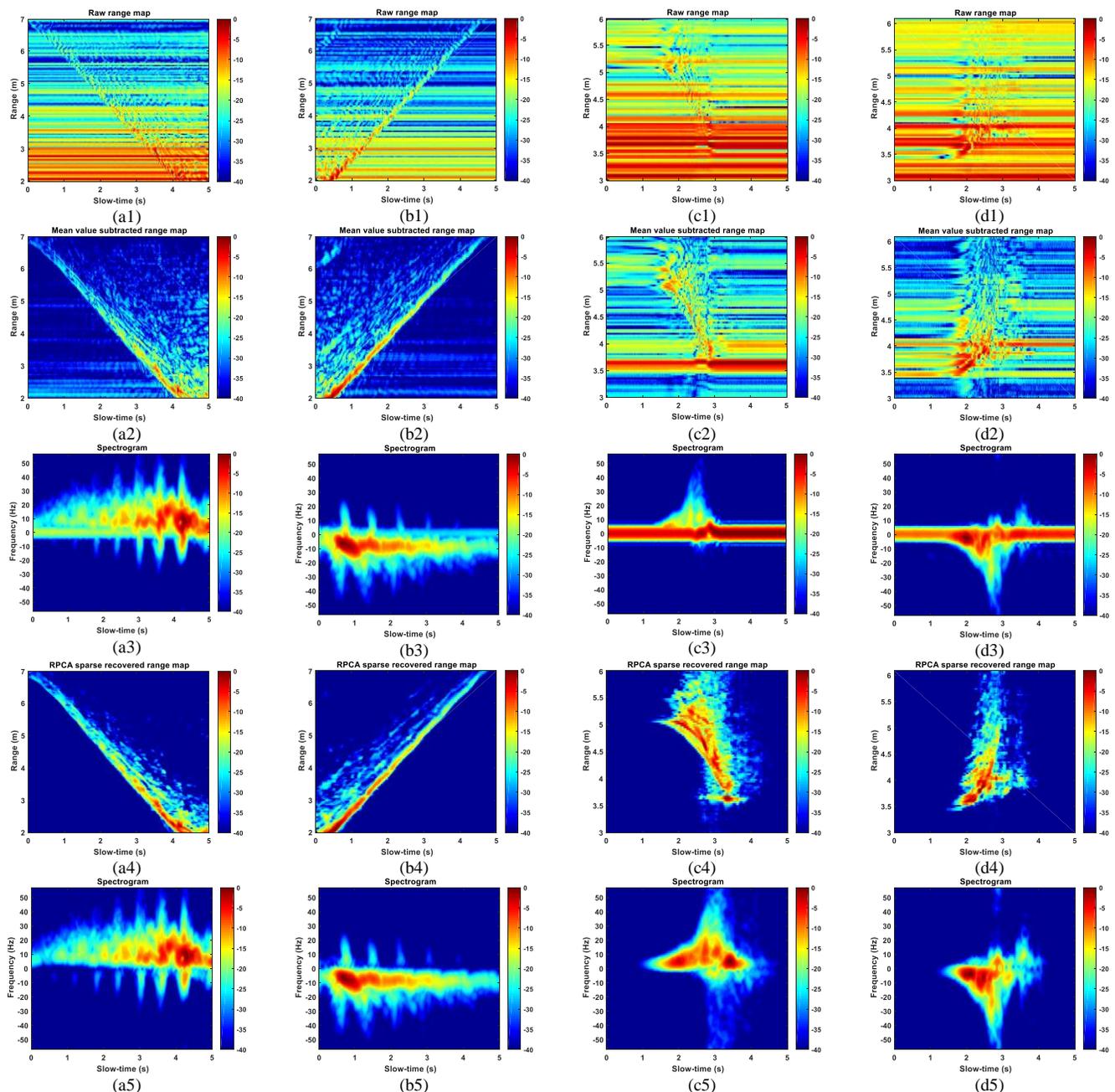

Figure 5: Through wall human motion detection results for four types of LRSM, namely, forward walking, backward walking, forward falling and backward falling. (a1-d1) Raw range maps. (a2-d2) Mean value subtraction filtered range maps. (a3-d3) Spectrograms of results in (a2-d2). (a4-d4) RPCA sparse reconstructed range map. (a5-d5) Spectrograms of results in (a4-d4).

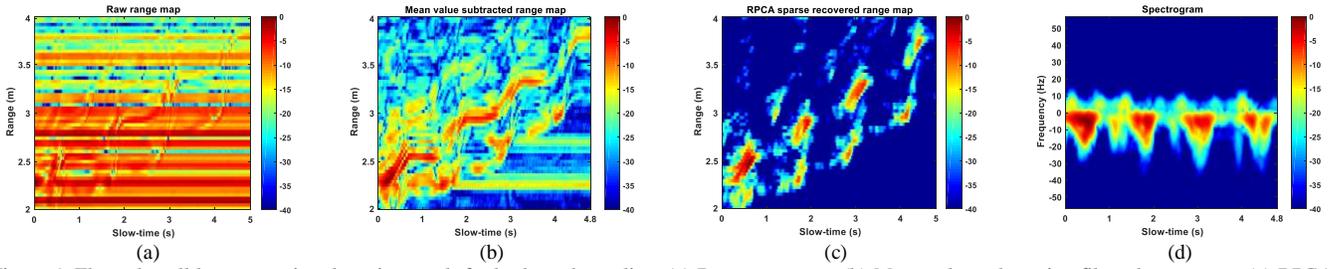

Figure 6: Through wall human motion detection result for backward crawling. (a) Raw range map. (b) Mean value subtraction filtered range map. (c) RPCA sparse reconstructed range map. (d) Spectrogram of result in (c).

collecting the EM reflections from body parts.

A total of 11 types of common indoor human movements are studied, including forward walking (type 1), backward walking (type 2), forward crawling (type 3), backward crawling (type 4), forward falling (type 5), backward falling (type 6), sitting on a chair (type 7), standing up (type 8), picking things up (type 9), in place marching (similar to walking on a treadmill, type 10) and boxing (type 11). These motions can be divided into two major categories, large range spanning motions (LRSM, types 1-6) and small range spanning motions (SRSM, types 7-11). The antenna array is regarded as the phase reference point for these motions. The behind-the-wall motion information is then captured using the aforementioned SFCW MIMO UWB radar. To be more specific, six antenna elements are deployed uniformly in horizontal observation line with a spacing of 0.4 m. Two of which located at leftmost and rightmost of the aperture function as transmitters, while the remaining elements in the middle work as receivers. In this way, a total number of eight equivalent channels are formed.

*A. RPCA-based micro-Doppler analysis for human motion in through wall scenario*

Fig. 5 shows the through-the-wall human motion detection results for four types of LRSM, which respectively include forward walking, backward walking, forward falling and backward falling. The first row of Fig. 5 denotes the raw range maps for the four LRSM, in which strong DC clutters, which are caused by reflections from the wall media and static indoor objects, totally obscure the behind-the-wall human movement information. The second row denotes the mean value subtraction clutter removal results. Only DC clutters are suppressed, while severe noise components and additional introduced DC clutters still populate the results. The third row shows the spectrograms of mean value subtraction filtered range map. The time-frequency features are critically masked by zero-frequency DC components. The fourth row shows the RPCA sparse recovered range maps. Significant clutters and noise components suppression performance is arrived using the approach. Discriminable motion information of behind-the-wall human subject is clearly observed in the results. The fifth row is the corresponding spectrograms. DC components in spectrograms are thoroughly removed when compared to the results in the third row. Simultaneously, the time-frequency features of different motions become easily distinguishable. For through-the-wall indoor motion monitoring application, one is specifically interested in the micro-Doppler signatures of two typical human motions when it comes to elderly care, namely forward falling and backward falling. The last two columns of the last row in Fig. 5 show exactly their respective micro-Doppler signatures, in which the characteristic 'positive tornado' and 'negative tornado' are sighted as indicated in millimeter-wave based fall detection [21]. This result further demonstrates the capability of the proposed technique for through-the-wall indoor human motion detection.

Fig. 6 shows the micro-Doppler signatures for the last type of LRSM, which is backward crawling. The positive frequency components in Fig. 6-(d) indicate the leg kick when a human subject crawls away from the antenna aperture. However, we do not see a similar negative equivalence for forward crawling in Fig. 3-(b2), which is due to the fact that radar cannot capture the leg kick motion when the human subject crawls towards the wall since the motion is weak in far distance than the forward movement of head and arms.

Fig. 7 denotes the through-the-wall micro-Doppler analysis results for four types of SRSM, which include respectively standing up, picking things up, in place marching and boxing. When compared to the results in Fig. 5, a similar conclusion can be readily drawn. Specifically, micro-Doppler signatures for sitting and standing up exhibit symmetrical features in Fig. 3-(a2) and Fig. 7-(a5). The two motions can then be easily discriminated from each other. The micro-Doppler signatures for picking things up is characterized by a small interval with apparently symmetrical features on the left and right in its own time-frequency map, as shown in Fig. 7-(b5). For in place marching motion, two split negative frequencies are representative features indicating respectively the arm pushed backward and the leg stepping down on the ground in Fig. 7-(c5). In addition, the amplitude for the negative and positive frequency is close. However, for the boxing motion, the positive frequency is a slightly larger than the negative one in Fig. 7-(d5). At the same time, no split of negative frequency is observed for this motion. We then can conclude that, even for the class of SRSM, RPCA works perfect in enhancing features of motions in both range map and time-frequency map. Thus, the 11 types of common indoor human motions discussed in this paper can be easily recognized using the proposed RPCA based feature enhancement method.

*B. Classification results*

In this part, we continue to investigate whether the above features avail to increasing the classification accuracy of different types of behind the wall indoor motions.

A matrix-based 2D-PCA dimensionality reduction method was first employed to do feature extraction in [22-24]. In comparison to standard PCA [25], [26], the matrix-based subspace technique does not require a matrix-to-vector conversion to compute the covariance matrices. Therefore, it can reduce the computational cost significantly. Then, in the classification stage, the k-Nearest Neighbors (kNN) classifier is employed to discriminate different motions. The key concept of kNN is to determine the label for each test sample by calculating its distance to $k$ closest training samples [27]. The value for $k$ is set to 3 in this work.

In order to verify the effectiveness of the RPCA based feature enhancement method, we conduct four separate classifications. The 2D-PCA based classification process consists of two stages, a training stage and a testing stage. In general, we follow the detailed classification steps given in [26] for standard 1-D PCA. These steps are omitted here for the sake of brevity. The mean-value-subtraction filtered range maps, its corresponding spectrograms, RPCA sparse recovered range maps and its corresponding spectrograms are utilized as inputs of classifier, respectively. The four datasets contain 9 classes of the above 11 motions, with in place marching and boxing are excluded since they don't belong to normal daily activities. For each motion, we have a total number of 320 samples. 80% of the dataset is used for training the classifier, while the rest of the dataset is used as testing. The training and testing sets are selected randomly. As has already pointed out in [21] and [26], the classification accuracy does not exhibit much improvement as one increases of the number of selected principal components. In this work, based on the observed dataset, sixteen dominant principal components are selected for the classification of mean value subtraction filtered range maps and RPCA sparse recovered range maps, whereas five principal components are used for their corresponding spectrograms.

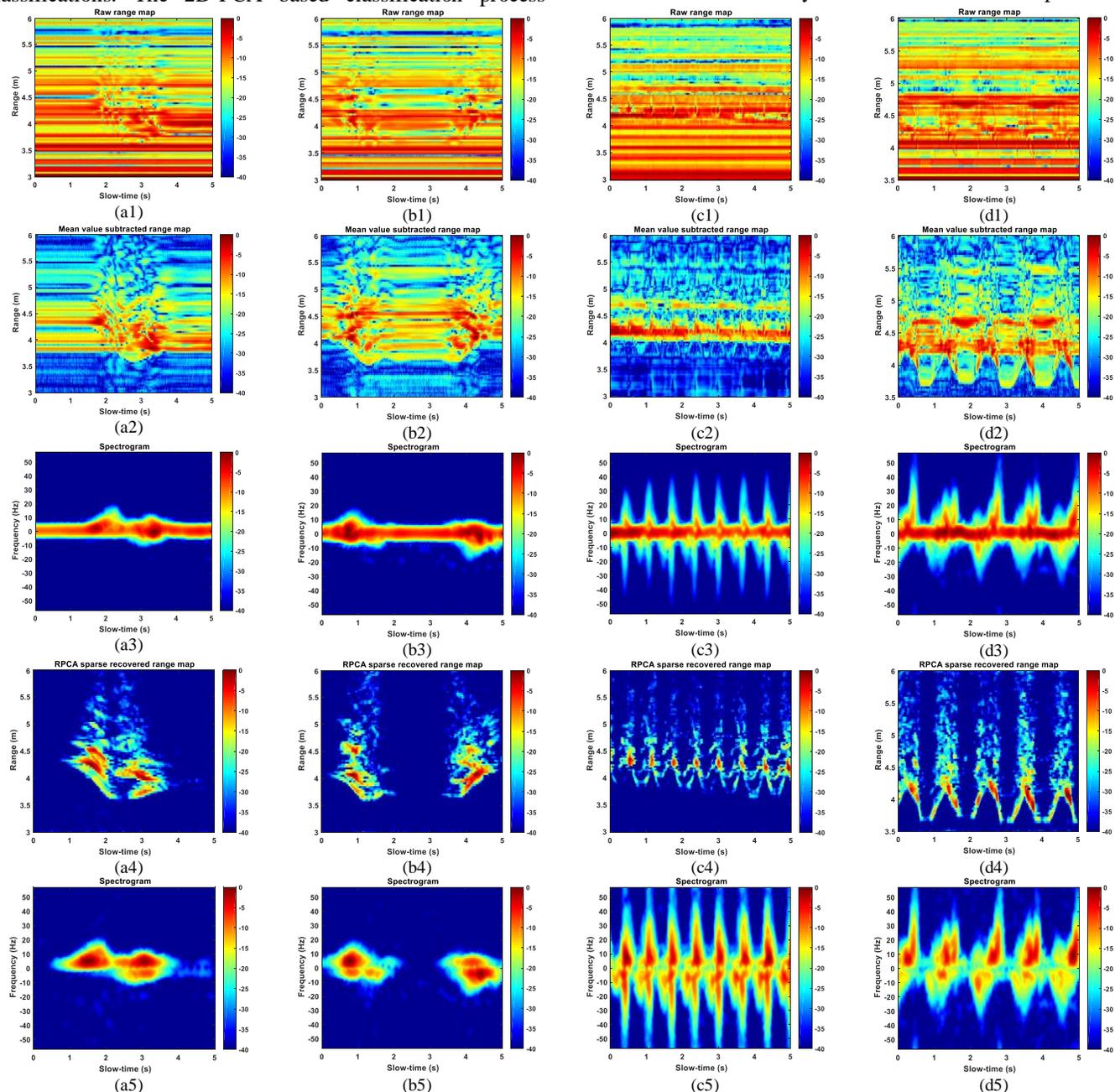

Figure 7: Through wall human motion detection results for four types of SRSM, namely, standing up, picking things up, in place marching and boxing. (a1-d1) Raw range maps. (a2-d2) Mean value subtraction filtered range maps. (a3-d3) Spectrograms of results in (a2-d2). (a4-d4) RPCA sparse reconstructed range map. (a5-d5) Spectrograms of results in (a4-d4).

TABLE I
Confusion matrix for mean value substracted range maps

|  | walk forward | walk backward | sit down | stand up | fetching | crawl forward | crawl backward | fall forward | fall backward |
|---|---|---|---|---|---|---|---|---|---|
| walk forward | 0.81 | 0.19 | 0.00 | 0.00 | 0.00 | 0.00 | 0.00 | 0.00 | 0.00 |
| walk backward | 0.17 | 0.83 | 0.00 | 0.00 | 0.00 | 0.00 | 0.00 | 0.00 | 0.00 |
| sit down | 0.00 | 0.00 | 0.84 | 0.15 | 0.00 | 0.00 | 0.01 | 0.00 | 0.00 |
| stand up | 0.00 | 0.00 | 0.08 | 0.89 | 0.01 | 0.00 | 0.01 | 0.01 | 0.00 |
| fetching | 0.00 | 0.00 | 0.00 | 0.00 | 0.99 | 0.01 | 0.00 | 0.00 | 0.00 |
| crawl forward | 0.00 | 0.00 | 0.00 | 0.02 | 0.05 | 0.88 | 0.05 | 0.00 | 0.00 |
| crawl backward | 0.00 | 0.00 | 0.00 | 0.00 | 0.00 | 0.02 | 0.98 | 0.00 | 0.00 |
| fall forward | 0.00 | 0.00 | 0.00 | 0.00 | 0.02 | 0.00 | 0.00 | 0.98 | 0.00 |
| fall backward | 0.00 | 0.00 | 0.00 | 0.02 | 0.00 | 0.09 | 0.00 | 0.00 | 0.89 |

TABLE II
Confusion matrix for spectrograms of mean value substracted range maps

|  | walk forward | walk backward | sit down | stand up | fetching | crawl forward | crawl backward | fall forward | fall backward |
|---|---|---|---|---|---|---|---|---|---|
| walk forward | 1.00 | 0.00 | 0.00 | 0.00 | 0.00 | 0.00 | 0.00 | 0.00 | 0.00 |
| walk backward | 0.00 | 1.00 | 0.00 | 0.00 | 0.00 | 0.00 | 0.00 | 0.00 | 0.00 |
| sit down | 0.00 | 0.00 | 0.96 | 0.01 | 0.02 | 0.00 | 0.01 | 0.00 | 0.00 |
| stand up | 0.00 | 0.00 | 0.00 | 0.95 | 0.02 | 0.02 | 0.00 | 0.01 | 0.00 |
| fetching | 0.00 | 0.00 | 0.02 | 0.00 | 0.98 | 0.00 | 0.00 | 0.00 | 0.00 |
| crawl forward | 0.00 | 0.00 | 0.02 | 0.00 | 0.00 | 0.94 | 0.00 | 0.04 | 0.00 |
| crawl backward | 0.00 | 0.00 | 0.01 | 0.00 | 0.00 | 0.00 | 0.92 | 0.00 | 0.07 |
| fall forward | 0.00 | 0.00 | 0.00 | 0.00 | 0.00 | 0.02 | 0.00 | 0.98 | 0.00 |
| fall backward | 0.00 | 0.00 | 0.00 | 0.00 | 0.00 | 0.00 | 0.02 | 0.00 | 0.98 |

TABLE III
Confusion matrix for RPCA sparse recovered range maps

|  | walk forward | walk backward | sit down | stand up | fetching | crawl forward | crawl backward | fall forward | fall backward |
|---|---|---|---|---|---|---|---|---|---|
| walk forward | 0.97 | 0.00 | 0.00 | 0.02 | 0.00 | 0.01 | 0.00 | 0.00 | 0.00 |
| walk backward | 0.00 | 1.00 | 0.00 | 0.00 | 0.00 | 0.00 | 0.00 | 0.00 | 0.00 |
| sit down | 0.00 | 0.02 | 0.95 | 0.03 | 0.00 | 0.00 | 0.00 | 0.00 | 0.00 |
| stand up | 0.00 | 0.00 | 0.00 | 0.98 | 0.00 | 0.02 | 0.00 | 0.00 | 0.00 |
| fetching | 0.00 | 0.00 | 0.01 | 0.00 | 0.99 | 0.00 | 0.00 | 0.00 | 0.00 |
| crawl forward | 0.00 | 0.00 | 0.00 | 0.02 | 0.00 | 0.98 | 0.00 | 0.00 | 0.00 |
| crawl backward | 0.00 | 0.00 | 0.00 | 0.00 | 0.01 | 0.00 | 0.99 | 0.00 | 0.00 |
| fall forward | 0.03 | 0.00 | 0.00 | 0.00 | 0.00 | 0.00 | 0.00 | 0.97 | 0.00 |
| fall backward | 0.00 | 0.00 | 0.00 | 0.00 | 0.00 | 0.00 | 0.02 | 0.00 | 0.97 |

TABLE IV
Confusion matrix for spectrograms of RPCA sparse recovered range maps

|  | walk forward | walk backward | sit down | stand up | fetching | crawl forward | crawl backward | fall forward | fall backward |
|---|---|---|---|---|---|---|---|---|---|
| walk forward | 1.00 | 0.00 | 0.00 | 0.00 | 0.00 | 0.00 | 0.00 | 0.00 | 0.00 |
| walk backward | 0.00 | 1.00 | 0.00 | 0.00 | 0.00 | 0.00 | 0.00 | 0.00 | 0.00 |
| sit down | 0.00 | 0.00 | 0.99 | 0.00 | 0.00 | 0.00 | 0.00 | 0.00 | 0.01 |
| stand up | 0.00 | 0.00 | 0.00 | 0.99 | 0.01 | 0.00 | 0.00 | 0.00 | 0.00 |
| fetching | 0.00 | 0.00 | 0.04 | 0.00 | 0.94 | 0.00 | 0.00 | 0.00 | 0.02 |
| crawl forward | 0.00 | 0.00 | 0.00 | 0.00 | 0.00 | 1.00 | 0.00 | 0.00 | 0.00 |
| crawl backward | 0.00 | 0.00 | 0.00 | 0.01 | 0.00 | 0.00 | 0.99 | 0.00 | 0.00 |
| fall forward | 0.00 | 0.00 | 0.00 | 0.00 | 0.00 | 0.00 | 0.00 | 1.00 | 0.00 |
| fall backward | 0.00 | 0.00 | 0.00 | 0.00 | 0.00 | 0.00 | 0.00 | 0.00 | 1.00 |

The classification accuracy is then assessed by the confusion matrix, which is a commonly used as a performance evaluation metric. It refers to the ability of correctly detecting a motion. The confusion matrixes for the four datasets are shown in Table I, Table II, Table III and Table IV, respectively. The overall classification accuracy for Table I is around 89%, which is much smaller than that in Table II, Table III and Table IV. Table III shows a close classification accuracy as in Table II. The result emphasizes the necessity of performing RPCA based feature enhancement prior to classification. However, since the input spectrograms is characterized by strong zero-frequency DC component for Table II, it is not suggested to use such feature maps for classification. In Table III, 97% of actual forward falling is classified as forward falling, while the rest (3%) is declared as forward walking. Similarly, only 2% of actual backward falling is classified as forward falling. The reason for such high classification accuracy lies in two aspects. One is that only a very limited number of observation angles is available in our experiments. The other is that 2-D PCA has been proved to yield higher classification accuracy than 1-D PCA in [29], which is widely used in millimeter wave based fall motion detections [7], [9], [21] and [26]. The results in Table IV show a similar accuracy as in Table III, which implies that both the RPCA sparse recovered range maps and the corresponding spectrograms can be utilized as feature maps for motion recognition and classification.

## V. CONCLUSION

In this paper, we present a RPCA-based behind-the-wall indoor human motion feature enhancement approach. The approach suppresses the stationary DC clutters, multipath and noise components simultaneously in the range map. The recovered sparse range map is clean with highlighted human subject movement trajectory. A DC clutter component removed time-frequency map with enhanced characteristic features is then obtained. Onsite experiment data of 11 behind-wall indoor motions is used to evaluate the proposed approach. 2-D PCA based classifications further demonstrate that the approach outperforms the traditional mean value subtraction based approach in behind-the-wall indoor human motion recognition.